\documentclass[a4paper,11pt]{article}
\usepackage{pos}

\title{Perspective on properties of renormalization schemes at high loops}
\ShortTitle{Perspective on renormalization schemes}

\author*[a]{J.A. Gracey}

\affiliation[a]{ Theoretical Physics Division,
Department of Mathematical Sciences,
University of Liverpool, P.O. Box 147, Liverpool, L69 3BX,
United Kingdom \\}

\emailAdd{gracey@liverpool.ac.uk}

\abstract{We report on recent work on a class of renormalization schemes in
QCD, termed the $\MOMts$ schemes. None of the renormalization group functions 
of the eight QCD $\MOMts$ schemes involve even zetas to five loops. A new 
scheme is introduced for scalar $\phi^3$ theory in six dimensions where the 
full Laurent series in the regularizing parameter of the Green's function is 
absorbed into the renormalization constants. Designated as the $\MaxSs$ scheme 
it is equivalent to the $\MOMts$ scheme in the critical dimension.}

\FullConference{Loops and Legs in Quantum Field Theory (LL2024)\\
14-19, April, 2024\\
Wittenberg, Germany\\}

\renewcommand{\logo}{\relax}

\newcommand{\MSbar}{\overline{\mbox{MS}}}
\newcommand{\MSbars}{\overline{\mbox{\footnotesize{MS}}}}
\newcommand{\MSbarss}{\overline{\mbox{\scriptsize{MS}}}}

\newcommand{\MOM}{{\mbox{MOM}}}

\newcommand{\MOMts}{\widetilde{\mbox{\footnotesize{MOM}}}}

\newcommand{\MOMtgggzgs}{\widetilde{\mbox{\footnotesize{MOM}}}_{ggg0g}}

\newcommand{\MOMtgggzggs}{\widetilde{\mbox{\footnotesize{MOM}}}_{ggg0gg}}

\newcommand{\MaxSs}{\overline{\mbox{\footnotesize{MaxS}}}}
\newcommand{\MaxSss}{\overline{\mbox{\scriptsize{MaxS}}}}
\newcommand{\iMOM}{\mbox{iMOM}}

\newcommand{\NA}{N_{\!A}}

\begin{document}
\vspace{-16.6cm}
\hspace{12.5cm}
{\bf LTH 1374}

\maketitle

\section{Introduction}

Over the last decade or so there have been signficant developments in the high 
loop order renormalization of gauge theories. For instance the $\MSbar$ 
$\beta$-function of Quantum Chromodynamics (QCD) is known to high precision, 
\cite{1,2,3,4,5,6,7,8,9,10}. Results for the renormalization group functions in
other schemes such as kinematic ones are not available to as many loops. For 
instance the QCD $\beta$-function in the momentum subtraction (MOM) schemes of 
Celmaster and Gonsalves, \cite{11}, is only available at four loops for the 
Landau gauge, \cite{12,13}. Once these core quantities are known the next stage
is to determine the perturbative expansion of observables to the same order of 
precision. In this respect while such quantities are invariably evaluated in 
the $\MSbar$ scheme there is no a priori reason why this scheme should be 
preferred over any other. They can equally well be determined in a MOM scheme 
for instance. In either case the observable will be available to a finite order
in the coupling constant expansion and therefore would only be an approximation
to the true value at some momentum scale. If one had a high enough number of 
terms then the theory uncertainty for a experimental measurement ought to be 
insignificant. Then the question arises as to how to arrive at an uncertainty 
value for the perturbative series truncation. One idea is to use the 
discrepancy in the value of the perturbative series when determined in several 
different schemes. Indeed an exploratory study of this idea was provided 
recently in \cite{14} for the R ratio and the Bjorken sum rule. For example 
using experimental data for the former quantity in the $\MSbar$, the MOM 
schemes of \cite{11} as well as the mini-MOM scheme of \cite{15}, estimates 
from the three and four loop expressions were extracted for 
$\alpha_s^{\MSbars}(M_Z)$, \cite{14}. These respectively were 
$0.13281 \pm 0.00197^{ +0.01171 }_{-0.00986}$ and
$0.13185 \pm 0.00053^{ +0.01072 }_{-0.00999}$. Here the error on the average is
the envelope of the scheme values and the average value is the centre of the 
envelope, \cite{14}. We record that these estimates are for an idealized 
situation where resonances and quark mass effects have not been taken into
account. The exercise was carried out by ignoring these aspects so that the 
scheme issue would be the sole focus of the study. It is reasonably apparent 
that the uncertainty reduces with increasing loop order. While this is 
encouraging what would be interesting is to include additional schemes in the 
analysis to check whether this improves the uncertainty in the sense of 
tightening it. We will review some recent activity in this direction for QCD 
here by discussing a suite of schemes introduced in \cite{16} and extended to 
five loops in \cite{17}.

\section{Basics}

We begin by recalling the basics behind defining a renormalization scheme. The 
Lagrangian is presented in terms of bare or classical variables that are not 
the optimum ones for describing quantum phenomena due to the presence of
infinities. Therefore the variables have to be redefined in terms of
renormalized ones which will lead to predictions from the field theory that are
devoid of divergences. The procedure to enact this is far from unique. There is
a requirement that after renormalization previously divergent Green's functions
are finite. Two criteria are used to define a scheme. First for renormalizable 
theories the Green's functions that are divergent are evaluated at a specific 
momentum configuration using the regularized Lagrangian. Then the combination 
of renormalization constants associated with that function are defined by a 
specified method called a scheme. The most basic scheme is the minimal 
subtraction one where only the poles with respect to the regulator are removed 
by fixing the unknown terms of the relevant renormalization constants. There 
are other schemes that render the Green's functions finite.

This can be illustrated with a simple massless cubic theory with Lagrangian
\begin{equation}
L ~=~ \frac{1}{2} \left( \partial_\mu \phi \right)^2 ~+~ \frac{g}{6} \phi^3
\label{lagphi3}
\end{equation}
where the coupling constant is $g$ and the critical dimension is six. In terms 
of bare variables the divergent Green's functions are
\begin{equation}
\Gamma_2(p) ~=~ \langle \phi_0(p) \phi_0(-p) \rangle ~~,~~
\Gamma_3(p_1,p_2,p_3) ~=~ \langle \phi_0(p_1) \phi_0(p_2) \phi_0(p_3) 
\rangle ~.
\end{equation}
To illustrate the structure of the Green's functions after renormalization we
note that for $\Gamma_2(p)$ it will take the two following forms
\begin{equation}
\left. \Gamma_2(p,-p) \right|_{p^2=\mu^2} ~=~ \left\{
\begin{array}{ll}
\mu^2 \left[ 1 ~+~ \sum_{n=1}^\infty a_n g^{2n} \right] & \MSbar \\
\mu^2 & \MOM \\
\end{array}
\right.
\end{equation}
after renormalization in the $\MSbar$ and MOM schemes respectively where $a_n$ 
are finite contributions. Here MOM denotes the momentum subtraction scheme of 
\cite{11} which has the prescription that the Green's function takes its tree 
value at the subtraction point. For the vertex function there are many more 
potential schemes given the larger choice of subtraction points. For example 
one can nullify an external momentum, which is infrared safe in six dimensions 
despite being an exceptional configuration, which introduces the $\MOMts$ 
schemes of \cite{16} defined by
\begin{equation}
\left. \Gamma_3(p,-p,0) \right|_{p^2=\mu^2} ~=~ \left\{
\begin{array}{ll}
g ~+~ \sum_{n=1}^\infty b_n g^{2n+1} & \MSbar \\
g & \MOMts ~. \\
\end{array}
\right. 
\end{equation}
Equally there are schemes for non-exceptional configurations one of which is
the symmetric point one considered in \cite{11} and defined by
\begin{equation}
\left. \Gamma_3(p_1,p_2,-p_1-p_2) \right|_{p_i^2=\mu^2} ~=~ \left\{
\begin{array}{ll}
g ~+~ \sum_{n=1}^\infty c_n g^{2n+1} & \MSbar \\
g & \MOM \\
\end{array}
\right.
\label{phi3momdef}
\end{equation}
for $i$~$=$~$1$ and $2$ with $p_3^2$~$=$~$\mu^2$ and $c_n$ are constants like 
$b_n$ but different in value since the subtraction points are not equivalent. 
The MOM scheme given in (\ref{phi3momdef}) is not unique since others can be 
constructed using different momentum configurations. For instance the 
interpolating momentum ($\iMOM$) subtraction scheme of \cite{18} defined by
\begin{equation}
\left. \Gamma_3(p_1,p_2,p_3) \right|_{p_1^2=p_2^2=\mu^2,p_3^2=\omega\mu^2} ~=~
\left\{
\begin{array}{ll}
g ~+~ \sum_{n=1}^\infty d_n g^{2n+1} & \MSbar \\
g & \iMOM \\
\end{array}
\right.
\end{equation}
is one such scheme that depends on the parameter $\omega$ defined for instance
by $\omega$~$=$~$\frac{p_3^2}{p_1^2}$~$=$~$\frac{p_3^2}{p_2^2}$ where the
$\omega$~$\to$~$1$ limit recovers the symmetric point of \cite{11}. However the
most general situation for a $3$-point vertex would regard these two 
dimensionless momenta ratios as independent. Given the wide variety of ways the
coupling can be renormalized opens up the possibility of having hybrid schemes. 
These can be constructed where for example the $2$-point function is rendered 
finite in $\MSbar$ say but the vertex function is made finite in one of the 
momentum subtraction schemes or its generalization to a $2$-variable scheme. 
Finally it is worth noting that the $\beta$-function of each of the schemes 
such as MOM, $\MOMts$ and $\iMOM$ carries information about the subtraction 
point itself via the finite parts that are related to $a_n$, $b_n$, $c_n$ and 
$d_n$. This first becomes evident at three loops in a single coupling theory or
two loops in a gauge theory for a non-zero covariant gauge parameter which 
produces a $\beta$-function which is gauge parameter dependent.

\section{$\MOMts$ examples}

To illustrate some of the properties of the $\MOMts$ scheme developed in
\cite{16} for QCD we have calculated the renormalization group functions in 
that scheme for $\phi^3$ theory to five loops. Two components are required to 
achieve this. One is the explicit form of the five loop $\MSbar$ 
renormalization group functions, which are already available in \cite{19,20}, 
and the other is the same quantities in the $\MOMts$ scheme as well as their 
corresponding four loop renormalization constants. The latter were computed
recently in \cite{21} using the {\sc Forcer} package provided in \cite{22}
written in {\sc Form}, \cite{23}. To use {\sc Forcer} for six dimensional 
computations required the determination of the {\sc Forcer} masters to high 
order in the $\epsilon$ expansion in $d$~$=$~$6$~$-$~$2\epsilon$ dimensions as 
(\ref{lagphi3}) is dimensionally regularized. The masters were deduced via the 
Tarasov method, \cite{24,25}, and provided in \cite{21} up to weight $9$ to be 
on an equivalent level to the four dimensional ones of the original package,
\cite{22}. The five loop $\MOMts$ renormalization group functions were then 
deduced via properties of the renormalization group equation. For instance the 
couplings in the respective schemes are related by
\begin{equation}
g_{{\MOMts}}(\mu) ~=~ \frac{Z_g^{\MSbarss}}{Z_g^{{\MOMts}}}
g_{{\MSbarss}}(\mu)
\label{ccmap}
\end{equation}
where
\begin{equation}
Z_g^{{\MOMts}} ~\equiv~ Z_g^{{\MOMts}}
\left( a_{\MOMts}(a_{\MSbarss}) \right)
\end{equation}
and $a$~$=$~$g^2$. Then the $\beta$-functions are related by
\begin{equation}
\beta_{\MOMts}^{\phi^3} ( a_{\MOMts} ) ~=~
\left[ \beta_{\mbox{$\MSbarss$}}^{\phi^3}( a_{\mbox{$\MSbarss$}} )
\frac{\partial a_{\MOMts}}{\partial a_{\mbox{$\MSbarss$}}}
\right]_{ \MSbarss \rightarrow {\MOMts} }
\end{equation}
where the mapping on the right hand side means the $\MSbar$ coupling is 
replaced by the inverse relation to (\ref{ccmap}). The field anomalous 
dimension can be deduced by a similar equation.

Consequently we have, \cite{21},
\begin{eqnarray}
\beta_{\MOMts}^{\phi^3}(a) &=&
\frac{3}{4} a^2 - \frac{125}{144} a^3
+ [ - 1296 \zeta_3 + 26741 ] \frac{a^4}{10368}
\nonumber \\
&&
+~ [ - 1370736 \zeta_3 + 2177280 \zeta_5 - 2304049 ] \frac{a^5}{186624}
\nonumber \\
&&
+~ [ 389670912 \zeta_3^2 + 3307195440 \zeta_3 + 89151840 \zeta_5
\nonumber \\
&& ~~~~~
- 5640570432 \zeta_7
+ 2190456157 ] \frac{a^6}{26873856} ~+~ O(a^7) 
\end{eqnarray}
where $\zeta_n$ is the Riemann zeta function. One interesting property is 
manifest and that is the absence of the even zetas, $\zeta_4$ and $\zeta_6$, 
which are present in the $\MSbar$ scheme $\beta$-function. Indeed this property
is not restricted to (\ref{lagphi3}) as it has been noted previously in QCD in 
\cite{16,26,27,28,29} and more recently checked for the core renormalization 
group functions, \cite{17}. More generally the criteria for the absence of 
$\pi$ in $\beta$-functions was formulated in the no-$\pi$ theorem, 
\cite{28,29}, having been motivated by observations in \cite{30}, and also 
discussed more recently in \cite{31} in the multicoupling context.

One application of the six dimensional {\sc Forcer} masters is to use them to
explore a generalization of the $\MOMts$ scheme definition. That scheme removed
the finite part of the $2$- and $3$-point functions at the subtraction point,
where one leg of the latter had a nullified momentum, in addition to the poles 
in $\epsilon$. This can be modified to the case where all the higher order 
powers in the $\epsilon$ expansion are removed too with the scheme being
designated the maximal subtraction scheme and labelled by $\MaxSs$, \cite{21}. 
The result of renormalizing $\phi^3$ theory in this scheme is to produce 
renormalization group functions in six dimensions which are formally equivalent
to those of the $\MOMts$ scheme. However this is not the case in the 
regularized theory where $\epsilon$~$\neq$~$0$ when 
$\beta_{\MaxSss}^{\phi^3}(a)$~$\neq$~$\beta_{\MOMts}^{\phi^3}(a)$ since the 
coefficient of the $O(\epsilon^n)$ term of a renormalization group function is 
related to the coefficient of the $O(\epsilon^{n-1})$ term of the corresponding
renormalization constant for $n$~$\geq$~$1$. This $\epsilon$ dependence in the 
$\beta$-function and other renormalization group functions is necessary to 
ensure that the critical exponent $\hat{\omega}$~$=$~$\beta^\prime(a^\star)$ is
the same in all schemes where $a^\star$ is the Wilson-Fisher fixed point in
$d$-dimensions. For (\ref{lagphi3}) we note that $\hat{\omega}$ only depends on
rationals and $\zeta_n$ for $3$~$\leq$~$n$~$\leq$~$5$ to $O(\epsilon^5)$.

The $\MOMts$ scheme is not restricted to (\ref{lagphi3}) and was actually
introduced for QCD in \cite{16} being developed originally at two loops. As QCD
has more than one cubic vertex with several involving different fields means 
there are several ways of nullifying external legs leading to quite a few
$\MOMts$ schemes. Subsequently the $\beta$-functions of several of these 
schemes were provided to four loops \cite{26,27}. More recently the 
renormalization group functions for eight possible $\MOMts$ schemes were 
determined to five loops in \cite{17}. The same renormalization group method as
discussed for $\phi^3$ theory above was used. In other words QCD was 
renormalized to four loops in the respective $\MOMts$ schemes using the bare 
$2$- and $3$-point functions provided in \cite{32} which were computed using 
the {\sc Forcer} package. Since the five loop QCD $\MSbar$ scheme 
renormalization group functions are available for an arbitrary colour group in 
\cite{32} all the ingredients are known to deduce the same data for the 
$\MOMts$ schemes of \cite{16}. For instance, the Yang-Mills (YM)
$\beta$-function for the $\MOMtgggzgs$ scheme is, \cite{17},
\begin{eqnarray}
\beta^{\mbox{\scriptsize{YM}}}_{\MOMtgggzgs} \!\!\!\!\!\!\!\!\!\! (a,0) &=&
-~ \frac{11}{3} C_A a^2
- \frac{34}{3} C_A^2 a^3
+ \left[
- \frac{6499}{48} C_A^3
+ \frac{253}{12} \zeta_3 C_A^3
\right] a^4
\nonumber \\
&&
+~ \left[
- \frac{10981313}{5184} C_A^4
- \frac{3707}{8} \zeta_3 \frac{d_A^{abcd} d_A^{abcd}}{\NA}
- \frac{8}{9} \frac{d_A^{abcd} d_A^{abcd}}{\NA}
\right. \nonumber \\
&& \left. ~~~~~~
+ \frac{6215}{24} \zeta_5 \frac{d_A^{abcd} d_A^{abcd}}{\NA}
+ \frac{97405}{576} \zeta_5 C_A^4
+ \frac{1116929}{1728} \zeta_3 C_A^4
\right] a^5
\nonumber \\
&&
+~ \left[
- \frac{8598255605}{165888} C_A^5
- \frac{1161130663}{73728} \zeta_7 C_A^5
\right. \nonumber \\
&& \left. ~~~~~~
- \frac{35208635}{3072} \zeta_7 C_A \frac{d_A^{abcd} d_A^{abcd}}{\NA}
- \frac{28905223}{2304} \zeta_3 C_A \frac{d_A^{abcd} d_A^{abcd}}{\NA}
\right. \nonumber \\
&& \left. ~~~~~~
- \frac{15922907}{9216} \zeta_3^2 C_A^5
+ \frac{131849}{3456} C_A \frac{d_A^{abcd} d_A^{abcd}}{\NA}
\right. \nonumber \\
&& \left. ~~~~~~
+ \frac{4595789}{384} \zeta_3^2 C_A \frac{d_A^{abcd} d_A^{abcd}}{\NA}
+ \frac{7284505}{1152} \zeta_5 C_A \frac{d_A^{abcd} d_A^{abcd}}{\NA}
\right. \nonumber \\
&& \left. ~~~~~~
+ \frac{30643529}{2048} \zeta_3 C_A^5
+ \frac{1667817635}{55296} \zeta_5 C_A^5
\right] a^6 ~+~ O(a^7)
\label{betamomtgggzg}
\end{eqnarray}
where the second argument of the $\beta$-function is the gauge parameter
$\alpha$, $\MOMtgggzgs$ denotes a $\MOMts$ scheme based on the triple gluon 
vertex, $C_A$, $C_F$, $T_F$ are the usual colour factors and
$d_A^{abcd} d_A^{abcd}$ is the rank four Casimir in the adjoint representation
of dimension $\NA$. Clearly (\ref{betamomtgggzg}) is devoid of even zetas as
are all the other five loop QCD $\MOMts$ renormalization group functions,
\cite{17}.

\section{Generalities}

Having provided an instance of a class of schemes with particular properties it
is worthwhile considering scheme changes from a more general perspective. First
we define the $\MSbar$ coupling renormalization constant and that of another
scheme ${\cal S}$ by
\begin{equation}
Z_g ~=~ 1 ~+~ \sum_{n=1}^\infty \sum_{m=1}^n z_{g\,nm}
\frac{a^n}{\epsilon^m} ~~~~,~~~~
Z_g^{\cal S} ~=~ 1 ~+~ \sum_{n=1}^\infty \sum_{m=0}^n z_{g\,nm}^{\cal S}
\frac{a_{\cal S}^n}{\epsilon^m}
\end{equation}
where there are finite contributions at each loop order in the scheme 
${\cal S}$. The respective couplings are perturbatively related by
\begin{equation}
a_{\cal S} ~=~ \sum_{n=0}^\infty c_n a^{n+1} ~.
\end{equation}
It is straightforward to show the connection the coefficients have with
$Z_g^{\cal S}$ with the few terms given by
\begin{eqnarray}
c_0 &=& 1 ~~~,~~~
c_1 ~=~ -~ 2 z_{g\,10}^{\cal S} ~~~,~~~
c_2 ~=~ 7 (z_{g\,10}^{\cal S})^2 ~-~ 2 z_{g\,20}^{\cal S} \nonumber \\
c_3 &=& -~ 30 (z_{g\,10}^{\cal S})^3 ~+~
18 z_{g\,10}^{\cal S} z_{g\,20}^{\cal S} ~-~
2 z_{g\,30}^{\cal S} 
\end{eqnarray}
illustrating that the $c_i$ depend solely on the finite parts of 
$Z_g^{\cal S}$. For instance the coupling constant mapping from the $\MSbar$ 
scheme to the $\MOMtgggzggs$ scheme in the Landau gauge is, \cite{17},
\begin{eqnarray}
a_{\MOMtgggzggs} &=& a
+ 16 a^2
+ \left[
\frac{93427}{192}
- \frac{169}{4} \zeta_3
\right] a^3
\nonumber \\
&&
+ \left[
\frac{129114635}{6912}
- \frac{1822913}{576} \zeta_3
- \frac{124835}{192} \zeta_5
\right] a^4
\nonumber \\
&&
+ \left[
\frac{4050665663}{4608}
- \frac{393488663}{2304} \zeta_3
+ \frac{980775}{512} \zeta_3^2
\right. \nonumber \\
&& \left. ~~~~~
+ \frac{1055749471}{36864} \zeta_7
- \frac{1387483355}{9216} \zeta_5
+ 1335 \zeta_4
\right] a^5 \,+\, O(a^6) 
\label{MOMtgggzggsccmap}
\end{eqnarray}
for the  $SU(3)$ colour group and three active quarks. While this contains 
$\zeta_4$ at four loops it is known that this term is key to the absence of 
$\zeta_4$ in the $\MOMtgggzggs$ renormalization group functions. The location
of the various $\zeta_n$ terms in (\ref{MOMtgggzggsccmap}) is mirrored in the
coupling constant maps for the other QCD $\MOMts$ schemes.

One reason for the $\MOMtgggzggs$ example rests in the potential connection
with the $C$-scheme of \cite{33}. That is a scheme which has its roots in the
relation of the $\Lambda$ parameters of two different schemes being related 
exactly by a one loop calculation \cite{11}. The relation depends on the one 
loop finite part of the coupling renormalization or $z_{g\,10}^{\cal S}$ for an
arbitrary scheme. In \cite{33} the four loop coupling constant map is provided 
for the $C$-scheme to $\MSbar$ for $SU(3)$ and three quark flavours. It too 
shares the property of the QCD $\MOMts$ scheme renormalization group functions 
in that no even zetas appear to five loops in various physical observables. So 
there is a possibility that one of the $\MOMts$ schemes could correspond to the 
$C$-scheme. The specifics of the $C$-scheme renormalization prescription have 
not been recorded. Instead only the coupling constant map has been provided,
\cite{33}, for three quarks. In particular, \cite{33},
\begin{equation}
a_C(a) ~=~ a ~-~ \frac{9}{4} C \, a^2 ~-~ \left[ \frac{3397}{2592} + 4C
- \frac{81}{16} C^2 \right] a^3 ~+~ O(a^4) 
\label{cschemeccmap}
\end{equation}
where $C$ is a free parameter within the $C$-scheme framework that can be tuned
to reduce uncertainties on observables. Its origin can be traced back to the
$\Lambda$ ratio between two schemes. In that respect it is akin to
$z_{g\,10}^{\cal S}$ and therefore we can use (\ref{cschemeccmap}) to see if a
connection can be made to one of the $\MOMts$ schemes. Examining
(\ref{MOMtgggzggsccmap}) we note that $\zeta_3$ appears at $O(a^3)$ but is not 
present in (\ref{cschemeccmap}) at the same order. A $\zeta_3$ could be
manufactured with a suitable choice of $C$ but that would mean $\zeta_3$ would
be present at $O(a^2)$. There are no such contributions in any of the QCD
$\MOMts$ scheme mappings at that order even when we consider those with a 
non-zero gauge parameter. So we believe the $C$-scheme does not correspond to
any of the $\MOMts$ schemes. However what all the $\MOMts$ schemes and 
$C$-scheme coupling mappings have in common is the $\zeta_4$ term at $O(a^5)$ 
in (\ref{MOMtgggzggsccmap}) with the same coefficient. It can be shown, 
\cite{17,28,29,31}, that this term is directly responsible for the absence of
$\zeta_4$ in the $\beta$-functions of these schemes.

We can also consider another general approach but in a different direction
which is to extend the $\iMOM$ scheme. Instead of a scheme depending on one
variable $\omega$ this can be replaced by the two independent parameters
\begin{equation}
x ~=~ \frac{p_1^2}{p_3^2} ~~~,~~~ y ~=~ \frac{p_2^2}{p_3^2}
\end{equation}
for $3$-point vertices. Consequently the renormalization functions will depend
on $x$ and $y$. For example the two loop Feynman gauge $SU(3)$ Yang-Mills 
$\beta$-function derived from the quark-gluon vertex is
\begin{eqnarray}
\left. \beta^{qqg}_{xy}(a,1) \right|_{\mbox{\scriptsize{YM}}}
&=& -~ 11 a^2 \nonumber \\
&& +~ \left[ [ 9 x^3 - 9 x^2 y - 54 x^2 - 9 x y^2 + 64 x y + 81 x + 9 y^3
\right. \nonumber \\
&& \left. ~~~~~~~
- 54 y^2 + 81 y - 36 ] \Phi_1(x,y) \Delta
- \frac{1064}{5} \Delta^2
\right. \nonumber \\
&& \left. ~~~~~
- [ 27 x^2 - 68 xy - 54 x + 41 y^2 - 68 xy + 27 ] \ln(xy) \Delta
\right] \frac{5a^3}{12\Delta^2}
\label{betaqqgxy}
\end{eqnarray}
where 
\begin{eqnarray}
\Phi_1(x,y) &=& \frac{1}{\lambda} \left[ 2 \mbox{Li}_2(-\rho x)
+ 2 \mbox{Li}_2(-\rho y)
+ \ln \left( \frac{y}{x} \right)
\ln \left( \frac{(1+\rho y)}{(1+\rho x)} \right)
\right. \nonumber \\
&& \left. ~~~~~
+ \ln(\rho x) \ln(\rho y) + \frac{\pi^2}{3} \right]
\end{eqnarray}
with
\begin{eqnarray}
\Delta(x,y) &=& x^2 ~-~ 2 x y ~+~ y^2 ~-~ 2 x ~-~ 2 y ~+~ 1 \nonumber \\
\lambda(x,y) &=& \sqrt{\Delta} ~~~,~~~
\rho(x,y) ~=~ \frac{2}{[1-x-y+\lambda(x,y)]} ~.
\end{eqnarray}
The $x$ and $y$ dependence in the two loop term does not contradict any general
properties since the $\beta$-function is gauge dependent. It is only in the
$\MSbar$ scheme of a single coupling theory that the $\beta$-function is scheme
independent to two loops. In schemes such as that which produced 
(\ref{betaqqgxy}) one can determine the perturbative expansion of observables 
as a function of $x$ and $y$. These parameters can then be varied to explore 
the uncertainty properties of the observable. In this respect $x$ and $y$ play 
a similar role to that of the free parameter $C$ of the $C$-scheme but have a
different origin. Moreover they could be restricted to a particular domain by
constraints on the kinematics.

\section{Conclusions}

One of the main observations is the renormalization group equations of both QCD
and scalar $\phi^3$ theory are free of even zetas up to five loops in the 
$\MOMts$ schemes. Moreover it has been shown that in the critical dimension of 
the latter theory the $\MaxSs$ and $\MOMts$ scheme renormalization group 
functions are equivalent. Indeed given the nature of both scheme definitions 
this property is probably applicable to theories other than $\phi^3$ theory. In
terms of usefulness of the results the availability of data on more schemes 
will mean that the estimate on the uncertainty deriving from the truncation of 
the perturbative expansion of an observable could in principle be improved. 
Finally while it is encouraging that the no-$\pi$ theorem of \cite{28,29} 
appears to hold to five loops in the $\MOMts$ schemes of \cite{16} it remains 
to be seen whether this continues to high loop order. There is evidence that 
the theorem may breakdown at very high loop order from the analysis of 
\cite{34}\footnote{We are grateful for David Broadhurst's comments on this 
point.}.

\acknowledgments
This work was carried out with the support of the STFC Consolidated Grant 
ST/T000988/1.
For the purpose of open access, the author has applied a Creative Commons 
Attribution (CC-BY) licence to any Author Accepted Manuscript version arising.

\end{document}